\def\Roman#1{\uppercase\expandafter{\romannumeral#1}}
\documentclass[a4paper,12pt]{article}
\usepackage{cite}
\usepackage{amssymb,amsfonts}
\usepackage[mathscr]{eucal}
\usepackage{mathrsfs}
\usepackage[utf8x]{inputenc}
\def\hyph{-\penalty0\hskip0pt\relax}

\title{Coordinate representation of 
the Lagrange-Poincar\'{e} equations for a mechanical system with symmetry on the total space of a principal fiber bundle whose base is    the bundle space of the associated bundle}

\author{S. N. Storchak\footnote{E-mail adress: storchak@ihep.ru}\\
\small{  NRC ``Kurchatov Institute'' -- IHEP,}\\
\small{Protvino, Moscow Region,  142281,   Russia}}

\begin{document}

  \maketitle

\begin{abstract}
Using the dependent coordinates, the local Lagrange-Poincaré equations and equations for the relative equilibria are obtained   for a mechanical system with a symmetry describing the motion of two interacting scalar particles on a special Riemannian  manifold (the product of  the total space of the principal fiber bundle and the vector space)   on which  a  free proper and isometric  action of a compact semi-simple Lie group is given. As in gauge theories, dependent coordinates are implicitly determined by means of  equations representing the local sections  of the principal fiber bundle.
\end{abstract}

\section{Introduction}

This note is a continuation of our previous work in which we obtained, in general form,  the local Lagrange-Poincaré equations (reduced Lagrange-Euler equations) for a finite-dimensional mechanical system describing the  two scalar particles with interaction  moving along a special Riemannian manifold represented by the product of two
manifolds $\mathcal P\times  V$,  on  which an action of a compact semi-simple group Lie $\mathcal G$ is given. The manifold 
$\mathcal P$, is a total space of the principal fiber bundle, and $ V$ is a finite-dimensional  vector space considered as manifold with an arbitrary constant metric.

The smooth,  free, proper, and isometric action action of a group $\mathcal G$ leads to the principal fiber bundle $ \mathcal P\times  V \to \mathcal P\times_{\mathcal G}  V $. And, therefore,  the original mechanical system can be reduced to the corresponding system given on the orbit space $\mathcal P\times_{\mathcal G}  V$ of this bundle.

In \cite{preprint} our goal was to obtain the local description of the evolution in the same way as it is done  in gauge field theories, i.e., using dependent variables. In gauge theories, these variables, implicitly defined, must satisfy  additional constraints (gauges) imposed on the gauge fields.   These constraints, given by equations,  determine the local gauge surface which is the local cross-section in the  principal fiber bundle associated  with the dynamical system. 
Using this surface, one can introduce  the local adapted coordinates in the prinipal fiber bundle. They are given by  dependent variables and group variables.

To introduce the adapted coordinates in the principal fiber bundle associated with our mechanical system, we used a local cross-section (the local "gauge surface") of that principal bundle for which the total space is a manifold $\mathcal P$.
How this can be done was shown in our previous work, where, using the Poincar\'{e} variational principle, we obtained  the Lagrange-Poincar\'{e} equations in general form.

In the present  notes, our goal is to get the coordinate representation for these equations. 
But before proceeding to this,  we first briefly recall the main points of our consideration performed in  our previous work.

\section{Coordinates on the configuration space}
As local  coordinates of a point $(p,v)$ given on our configuration space, the manifold  $\mathcal P \times V$, we take 
$(Q^A,f^n)$, $A=1,\dots , N_P$ and $n=1,\dots ,N_V$ such that $Q^A=\varphi ^A(p)$, and $f^n=\varphi ^n(v)$, where $(\varphi ^A,\varphi ^n)$ are the coordinate functions of  a chart on the original product manifold.

In these coordinates, the Riemannian metric of the manifold is written as follows:
\begin{equation}
 ds^2=G_{AB}(Q)dQ^AdQ^B+G_{mn}df^mdf^n.
\label{metr_orig}
\end{equation}
The  matrix $G_{mn}$ consists of some fixed constant  elements.

The right action of the group $\mathcal G$,
$(p,v)g=(pg,g^{-1}v)$, is 
 written  as 
\[
 {\tilde Q}^A=F^A(Q,g),\;\;\;\;{\tilde f}^n=\bar D^n_m(g)f^m.
\]
Here $\bar D^n_m(g)\equiv D^n_m(g^{-1})$,
and  $D^n_m(g)$ is the matrix of  the finite-dimensional representation of the group $\mathcal G$
acting on the vector space $V$.

Due to isometry, we have two relations for the metric tensors:
\begin{equation}
 G_{AB}(Q)=G_{DC}(F(Q,g))F^D_A(Q,g)F^C_B(Q,g),
\label{relat_G_AB}
\end{equation}
with $F^B_A(Q,g)\equiv\frac{\partial F^B(Q,g))}{\partial Q^A}$, and
\begin{equation}
 G_{pq}=G_{mn}\bar D^m_p(g)\bar D^n_q(g).
\label{relat_g_mn}
\end{equation}

The Killing vector fields are defined as \\
$K^A_{\alpha}(Q)\frac{\partial}{\partial Q^A}$, with
$K^A_{\alpha}(Q)=\frac{\partial {\tilde Q}^A}{\partial a^{\alpha}}\Big|_{a=e}$, and
$K^n_{\alpha}(f)\frac{\partial}{\partial f^n}$, with $K^n_{\alpha}(f)=\frac{\partial {\tilde f}^n}{\partial a^{\alpha}}\Big|_{a=e}=\frac{\partial {\bar D}^n_m(a)}{\partial a^{\alpha}}\Big|_{a=e}=({\bar J}_{\alpha})^n_m f^m$.  The generators ${\bar J}_{\alpha}$ of the representation ${\bar D}^n_m(a)$ have the following  commutation relation: 
$[{\bar J}_{\alpha},{\bar J}_{\beta}]={\bar c}^{\gamma}_{\alpha \beta}{\bar J}_{\gamma}$, where the structure constants
${\bar c}^{\gamma}_{\alpha \beta}=-{c}^{\gamma}_{\alpha \beta}$.

We use  the (condensed) notation by which the capital Latin letter with tilde represents two subscripts (or superscripts) that are  related with two spaces: 
$\tilde A\equiv (A,p)$. For  components of the Killing vector fields, for example, we have
$$ K^{\tilde A}_{\mu}=(K^{A}_{\mu},K^{p}_{\mu}).$$

From the general theory \cite{AbrMarsd} it follows that in our case we can regard the original manifold as a total space of the principal fiber bundle $$\pi': \mathcal P\times V\to \mathcal P\times _{\mathcal G}V,$$
where $\pi': (p,v)\to [p,v]$, and  $[p,v]$ is the equivalence class formed by the equivalence relation   $(p,v)\sim (pg,g^{-1}v)$.)  
This allows us to introduce new  coordinates on $\mathcal P\times V$ that are related with this principal fiber bundle. Moreover, one can  
 express the coordinates $(Q^A, f^n)$ of the point $(p,v)$ in terms of the  bundle coordinates by the  well-known procedure 
\cite{Creutz, Razumov, Storchak_11, Storchak_12, Storchak_2, Storchak_3, Huffel-Kelnhofer, Kelnhofer}.

In this procedure, the bundle coordinates are introduced with the help of the local section $\tilde \sigma_i$ of the bundle, $\pi' \cdot\tilde \sigma_i = \rm{id}$,  sending the point $[p,v]$ to some element $(\tilde p,\tilde v)\in \mathcal P\times V$.   $\tilde\sigma_i$ is defined as follows:
\[
 \tilde \sigma_i([p,v])=(\sigma_i(x),a(p) v), 
\]
where $\sigma_i$ is a local  section of the principal fiber bundle $\rm P(\mathcal M,\mathcal G)$ 
with the base space  $\mathcal M= \mathcal P/\mathcal G$,  
$\sigma_i:U_i\to\pi_{\rm P}^{-1}(U_i)$, $x=\pi_{\rm P}(p)$ and $a(p)$ is the group element defined by $p=\sigma_i(x)a(p)$.
Also note that due to 
\[
 (\sigma_i(x),a(p)\, v)=(p\,a^{-1}(p),a(p)\, v)=(p,v)\,a^{-1}(p),
\]
we have
\[
 \tilde \sigma_i([p,v])=(p,v)\,a^{-1}(p).
\]

The local sections $\sigma_i$ of the principal fiber bundle $\rm P(\mathcal M,\mathcal G)$  in a
neighbourhood of a point $p\in \mathcal P$ 
can be  determined by   local submanifolds   $\Sigma_{i}$ which have the transversal intersections with  the orbits.
The section $\sigma_i$ is  the map $\sigma_i:U_i\to \Sigma_i$ such that  $\pi_{\Sigma_i}\cdot\sigma_i={\rm id}_{U_i}$.
In its turn, these submanifolds are given by the equations $\{\chi^{\alpha}(Q)=0, \alpha= 1,\ldots N_{\mathcal G}\}$.

The  coordinates  of the points on the local submanifold $\Sigma_i$ 
will be denoted by $Q^{\ast}{}^A$. Since they satisfy the equations $\{\chi^{\alpha}(Q^{\ast})=0\}$, they are called the dependent coordinates.  Any point   $p$  on the total space $\mathcal P$ of the principal fiber bundle $\rm P(\mathcal M,\mathcal G)$ 
must have, in addition, a group coordinates $a^{\alpha}$. 

 A local isomorphism between trivial principal bundle $\Sigma_{i}\times \mathcal G\to\Sigma_{i}$ and 
$\rm P(\mathcal M,\mathcal G)$\cite{Babelon-Viallet, Huffel-Kelnhofer, Kelnhofer}  
\[
 \varphi_{i}:\,\,\Sigma_{i}\times \mathcal G\to \pi ^{-1}(U_{i})
\]
is given in coordinates as
\[
 \varphi_{i}:(Q^{\ast}{}^B,a^{\alpha})\to Q^A=F^A(Q^{\ast}{}^B, a^{\alpha}),
\]
where $Q^{\ast}{}^B$ are the coordinates of a  point given  on the local surface $ \Sigma_{i}$ and 
$a^{\alpha}$ -- the coordinates of an arbitrary group element $a$. This element carries the point, taken on $ \Sigma_{i}$, to  the point $p\in \mathcal P$ which has  the coordinates $Q^A$.

An inverse map $\varphi_{i}^{-1}$,
\[
 \varphi_{i}^{-1}:\,\,\pi ^{-1}(U_{i})\to\Sigma_{i}\times \mathcal G,
\]
has the following coordinate representation:
\[
 \varphi_{i}^{-1}: Q^A\to (Q^{\ast}{}^B(Q),a^{\alpha}(Q)).
\]
Here the group coordinates $a^{\alpha}(Q)$ of a point $p$ are the coordinates of the group element  
which connects, by means of its  action on $p$, the surface $\Sigma _{i}$ and the point $p\in \mathcal P$. These group coordinates are given by the solutions of the following equation:
\begin{equation}
 \chi^{\beta}(F^A(Q, a^{-1}(Q)))=0.
\label{a_chi}
\end{equation} 
The coodinates $Q^{\ast}{}^B$ are defined by the equation
\begin{equation}
 Q^{\ast}{}^B=F^B(Q, a^{-1}(Q)).
\label{Q_star}
\end{equation}

In the same way as for the principal bundle $\rm P(\mathcal M,\mathcal G)$, there exist a local isomorphisms of the principal fiber bundle ${\rm P}(\mathcal P\times _{\mathcal G}V,\mathcal G)$ and the  trivial principal bundles $\tilde \Sigma_i\times \mathcal G \to \tilde \Sigma_i$, where now the local surfaces $\tilde \Sigma_i$ are the images of the sections $\tilde \sigma_i$.

Therefore, 
we can introduce a new atlas on ${\rm P}(\mathcal P\times _{\mathcal G}V,\mathcal G)$.
In this atlas, the coordinate functions  of the charts $(\tilde U_i,\tilde\varphi_i)$, where $\tilde U_i$ is an open neighborhood of the point $[p,v]$ given on the base space  $\mathcal P\times_{\mathcal G}V$, are such that
\[
 \tilde \varphi_i^{-1} : \pi^{-1}(\tilde U_i)\to\tilde \Sigma_i \times \mathcal G,\;\;{\rm or}\; {\rm  in}\; {\rm coordinates,} 
\]
\[
 \tilde \varphi_i^{-1} :(Q^A,f^m)\to (Q^{\ast}{}^A(Q),\tilde f^n(Q),a^{\alpha}(Q)\,).
\]
Here  $Q^A$ and $f^m$ are the coordinates of a point $(p,v)\in \mathcal P\times V$,  
$Q^{\ast}{}^A(Q)$ is given by (\ref{Q_star}) and
\[
\tilde f^n(Q) = D^n_m(a(Q))\,f^m,
\]
$a(Q)$ is defined by  (\ref{a_chi}), and  we have used the following property: $\bar D^n_m(a^{-1})\equiv D^n_m(a)$.  The coordinates $Q^{\ast}{}^A$, representing  a point given on a local surface $\Sigma_i$,  satisfy the constraints: $\chi(Q^{\ast})=0$. That is, they are dependent coordinates.

The coordinate function $\tilde \varphi_i$ maps $\tilde \Sigma_i\times \mathcal G\to \pi^{-1}(\tilde U_i)$:
\[
 \tilde \varphi_i :(Q^{\ast}{}^B,\tilde f^n,a^{\alpha})\to (F^A(Q^{\ast},a), \bar D^m_n(a)\tilde f^n).
\]
Thus, we have defined the special  local bundle coordinates $(Q^{\ast}{}^A,\tilde f^n, a^{\alpha})$, also named as  adapted coordinates, 
on the principal fiber bundle 
$\pi:\mathcal P\times V\to \mathcal P\times_{\mathcal G} V$.

It is not difficult to obtain the representation for the Riemannian metric  given on $\mathcal P\times V$ in terms of the principal bundle  coordinates $(Q^{\ast}{}^A,{\tilde f}^n, a^{\alpha})$. The replacement of the coordinates $(Q^A,f^m)$ of a point $(p,v)\in \mathcal P\times V$  for  new coordinates
\begin{equation}
Q^A=F^A(Q^{\ast}{}^B,a^{\alpha}),\;\;\;f^m=\bar D^m_n(a)\tilde f^n
\label{transf_coord}
\end{equation}
 leads to the following
transformation of the local coordinate vector fields: 
\begin{eqnarray}
\displaystyle 
&&\!\!\!\!\!\!\!\!\frac{\partial}{\partial f^n}=D^m_n(a)\frac{\partial}{\partial {\tilde f}^m},
\nonumber\\
&&\!\!\!\!\!\!\!\!\frac{\partial}{\partial Q^B}=\frac{\partial Q^{\ast}{}^A}{ \partial Q^B}\frac{\partial}{\partial Q^{\ast}{}^A}+\frac{\partial a^{\alpha}}{\partial Q^B}\frac{\partial}{\partial a^{\alpha}}+\frac{\partial {\tilde f}^n}{\partial Q^B}\frac{\partial}{\partial {\tilde f}^n}
\nonumber\\
&&\!\!\!\!\!\!\!\!\!\!\!=\check F^C_B\Biggl(N^A_C(Q^{\ast})\frac{\partial}{\partial Q^{\ast}{}^A}+{\chi}^{\mu}_C({\Phi}^{-1})^{\beta}_{\mu}\bar{v}^{\alpha}_{\beta}(a)\frac{\partial}{\partial a^{\alpha}}-{\chi}^{\mu}_C({\Phi}^{-1})^{\nu}_{\mu}(\bar J_{\nu})^m_p\tilde f^p\frac{\partial}{\partial {\tilde f}^m}\Biggr).\:
\label{vectfield}
\end{eqnarray}
Here $\check F^C_B\equiv F^C_B(F(Q^{\ast},a),a^{-1})$ is an inverse matrix to the matrix $F^A_B(Q^{\ast},a)$,
${\chi}^{\mu}_C\equiv \frac{\partial {\chi}^{\mu}(Q)}{\partial Q^C}|_{Q=Q^{\ast}}$, $({\Phi}^{-1})^{\beta}_{\mu}\equiv({\Phi}^{-1})^{\beta}_{\mu}(Q^{\ast})$ -- the matrix which is inverse to the Faddeev--Popov matrix:
\[
 ({\Phi})^{\beta}_{\mu}(Q)=K^A_{\mu}(Q)\frac{\partial {\chi}^{\beta}(Q)}{\partial Q^A},
\]
the matrix $\bar{v}^{\alpha}_{\beta}(a)$ is inverse of the matrix $\bar{u}^{\alpha}_{\beta}(a)$.\footnote{$\det \bar{u}^{\alpha}_{\beta}(a)$ is   the density of the right-invariant measure  given on the group $\mathcal G$.}

The operator $N^A_C$, defined  as
\[
 N^A_C(Q)=\delta^A_C-K^A_{\alpha}(Q)({\Phi}^{-1})^{\alpha}_{\mu}(Q){\chi}^{\mu}_C(Q), 
\]
 is the projection operator ($N^A_BN^B_C=N^A_C$) onto the subspace which is orthogonal to the Killing vector field $ K^A_{\alpha}(Q)\frac{\partial}{\partial Q^A}$. $N^A_C(Q^{\ast})$ is the restriction of $N^A_C(Q)$ to the submanifold $ \Sigma $:
\[
 N^A_C(Q^{\ast})\equiv N^A_C(F(Q^{\ast},e))\;\;\;N^A_C(Q^{\ast})=F^B_C(Q^{\ast},a)N^M_B(F(Q^{\ast},a))\check F_M^A(Q^{\ast},a)
\]
$e$ is the unity element of the group. 

We note also that  formula (\ref{vectfield}) is a generalization of  an analogous formula
from \cite{Storchak_11,Storchak_2}.

 The vector field $\frac{\partial}{\partial Q^{\ast}{}^A}$ is determined as an operator using the following rule:\footnote{It can be shown that this rule follows from the approaches developed in \cite{Creutz} and  \cite{Plyush}.}
\[
 \frac{\partial}{\partial Q^{\ast}{}^A }\varphi(Q^{\ast})=(P _\bot)^D_A(Q^{\ast})\frac{\partial \varphi(Q)}{\partial Q^D}\Bigl|_{Q=Q^{\ast}},
\]
where the projection operator $(P_\bot)^A_B$ on the tangent plane to the submanifold $\Sigma$ is given by
\[
 (P_\bot)^A_B=\delta^A_B-\chi ^{\alpha}_{B}\,(\chi \chi ^{\top})^{-1}{}^{\beta}_{\alpha}\,(\chi ^{\top})^A_{\beta}. 
\]
In this formula, $(\chi ^{\top})^A_{\beta}$ is a transposed matrix to the matrix $\chi ^{\nu}_B$:
\[
 (\chi ^{\top})^A_{\mu}=G^{AB}{\gamma}_{\mu\nu}\chi ^{\nu}_B\;\;\;\;{\gamma}_{\mu\nu}=K^A_{\mu}G_{AB}K^B_{\nu}.
\]
Using the above explicit expression
 for the projection operators, it is easy to derive 
their multiplication properties:
\[
 (P_\bot)^A_BN^C_A=(P_\bot)^C_B,\;\;\;\;\;N^A_B(P_\bot)^C_A=N^C_B.
\]

In a new coordinate basis 
$\displaystyle(\partial/\partial Q{}^{\ast A},\partial/\partial \tilde f^m,\partial/\partial a^{\alpha})$,
the metric (\ref{metr_orig}) of the original manifold $\mathcal P \times V$ is represented by means of the following tensor:
\begin{equation}
\displaystyle
{\tilde G}_{\cal A\cal B}(Q{}^{\ast},\tilde f,a)=
\left(
\begin{array}{ccc}
 G_{CD}(P_{\bot})^C_A (P_{\bot})^D_B & 0 & G_{CD}(P_{\bot})^C_AK^D_{\nu}\bar u^{\nu}_{\alpha}\\
 0 & G_{mn} & G_{mp}K^p_{\nu}\bar u^{\nu}_{\alpha}\\
G_{BC}K^C_{\mu}\bar u^{\mu}_{\beta} & G_{np}K^p_{\nu}\bar u^{\nu}_{\beta} & d_{\mu\nu}\bar u^{\mu}_{\alpha}\bar u^{\nu}_{\beta}\\
\end{array}
\right)
\label{metric2c}
\end{equation}
where $G_{CD}(Q{}^{\ast})\equiv G_{CD}(F(Q{}^{\ast},e))$:
\[
 G_{CD}(Q{}^{\ast})=F^M_C(Q{}^{\ast},a)F^N_D(Q{}^{\ast},a)G_{MN}(F(Q{}^{\ast},a)), 
\]
the projection operators $P_\bot$ and the components $K^A_{\mu}$ of the Killing vector fields  depend on $Q{}^{\ast}$, $\bar u^{\mu}_{\beta}=\bar u^{\mu}_{\beta}(a)$,  $K^p_{\nu}=K^p_{\nu}(\tilde f)$,  $d_{\mu\nu}(Q{}^{\ast},\tilde f)\bar u^{\mu}_{\alpha}(a)\bar u^{\nu}_{\beta}(a)$ is 
the metric on  $\mathcal G$--orbit through the point $(p,v)$. The components $d_{\mu\nu}$ of this metric are  given by
\begin{eqnarray*}
 d_{\mu\nu}(Q{}^{\ast},\tilde f)&=&K^A_{\mu}(Q{}^{\ast})G_{AB}(Q{}^{\ast})K^B_{\nu}(Q{}^{\ast})+K^m_{\mu}(\tilde f)G_{mn}K^n_{\nu}(\tilde f)
\nonumber\\
&\equiv&\gamma_{\mu \nu}(Q^{\ast})+\gamma'_{\mu \nu}(Q^{\ast}).
\end{eqnarray*}

\section{Transformation of the Lagrangian}

In terms of  initial  local coordinates  defined on the original manifold $\mathcal P\times V$, the  Lagrangian for the considered mechanical system can be written  as follows:
\begin{equation}  
\mathcal L=\frac12 G_{AB}(Q)\,{\dot Q}^A{\dot Q}^B +\frac12 G_{mn}\,{\dot f}^m{\dot f}^n-V(Q,f).
\label{lagrang_1}
\end{equation}
 By our assumption,  the potential $V(Q,f)$ is a $\mathcal G$-invariant function, that is,  $V(Q,f)=V(F(Q,a),\bar D(a)f)$. So the whole Lagrangian is also invariant.

 The replacement of the local coordinates  (\ref{transf_coord}), which introduces the coordinates  
$({Q^{\ast}}^{A},{\tilde f}^m, a^{\alpha})$
on $\mathcal P\times V$, transforms the Lagrangian into
\begin{eqnarray}
 &&{\mathcal L}=\frac12G_{CD}\Bigl(\frac{d{Q^{\ast}}^{C}}{dt}+K^C_{\mu}\,{\bar u}^{\mu}_{\alpha}(a)\,\frac{da^{\alpha}}{dt}\Bigr)\Bigl(\frac{d{Q^{\ast}}^{D}}{dt}+K^D_{\nu}\,{\bar u}^{\nu}_{\beta}(a)\,\frac{da^{\beta}}{dt}\Bigr)
\nonumber\\
&&\;\;\;\;\;+\frac12G_{mn}\Bigl(\frac{d{\tilde f^m}}{dt}+K^m_{\beta}\,{\bar u}^{\beta}_{\alpha}(a)\,\frac{da^{\alpha}}{dt}\Bigr)\Bigl(\frac{d{\tilde f^n}}{dt}+K^n_{\nu}\,{\bar u}^{\nu}_{\mu}(a)\,\frac{da^{\mu}}{dt}\Bigr)-V,
\label{lagrang_2}
\end{eqnarray}
where now $G_{CD}, K^C_{\mu}$ depend on $Q^{\ast}$, $K^m_{\beta}=K^m_{\beta}(\tilde f)$, and $V=V(Q^{\ast},\tilde f)$.

The Lagrange-Poincar\'e  equations are obtained with the help of 
 a special coordinate basis (the horizontal lift basis)  on the total space of the principal fiber bundle. 
The new basis consists of the horizonal and vertical vector fields and can be determined by using the ``mechanical connection'' which 
exists \cite{AbrMarsd} in  case of the reduction of  mechanical systems with a symmetry.

The connection one-form  $\hat\omega ^{\alpha}$   in the principal fiber bundle ${\rm P}(\mathcal P\times_{\mathcal G}V,\mathcal G)$,\footnote{The one-form $\hat\omega$ with the value in the Lie algebra of the group Lie $\mathcal G$ is  $\hat\omega=\hat\omega ^{\alpha}\otimes\lambda_{\alpha}$.}  is given by the following formula written in terms of the initial local coordinates defined  
on the total space $\mathcal P\times V$:
\begin{equation}
 \hat\omega ^{\alpha}(Q,f)=d^{\alpha\beta}(Q,f)\,(K^B_{\beta}(Q)G_{BA}(Q)dQ^A+K^p_{\beta}(f)G_{pq}df^q\,).
\label{connect_Q}
\end{equation}
In coordinates $(Q^{\ast}{}^A,\tilde f^n,a^{\alpha})$, the one-form is written as follows:
\[
 \hat\omega ^{\alpha}=\bar {\rho}^{\alpha}_{\alpha '}(a)\biggl(d^{\alpha '\mu}K^D_{\mu}(Q{}^{\ast})G_{DA}(Q{}^{\ast})dQ{}^{\ast  A}+d^{\alpha '\mu}K^q_{\mu}(\tilde f)G_{qn}d{\tilde f}^n\biggr)+u^{\alpha}_{\beta}(a)da^{\alpha},
\]
where now  $d^{\alpha '\mu}= d^{\alpha '\mu}(Q{}^{\ast},\tilde f)$. And the  matrix  $\bar {\rho}^{\alpha}_{\alpha '}(a)$ 
 is inverse to the matrix ${\rho}_{\alpha}^{\beta}={\bar u}^{\alpha}_{\nu}v^{\nu}_{\beta}$  of the adjoint representation of the group $\cal G$,

Introducing the (gauge) potentials $\mathscr A^{\alpha }_B$, and $\mathscr A^{\alpha '}_m$, together with  a new notation: $\tilde\mathscr A^{\alpha }_B=\bar {\rho}^{\alpha}_{\alpha '}(a)\mathscr A^{\alpha '}_B(Q{}^{\ast},\tilde f)$, we come to
\begin{equation}
\hat\omega ^{\alpha}= 
\tilde\mathscr A^{\alpha '}_B(Q{}^{\ast},\tilde f,a)dQ{}^{\ast  B}+\tilde\mathscr A^{\alpha '}_m(Q{}^{\ast},\tilde f,a)d{\tilde f}^m+u^{\alpha}_{\beta}(a)da^{\alpha}.
\label{connect_Q_star}
\end{equation}
In term of  condensed notations of  indices, $\hat\omega ^{\alpha}$ is written as
\[
 \hat\omega ^{\alpha}= 
\tilde\mathscr A^{\alpha '}_{\tilde B}(Q{}^{\ast},\tilde f,a)dQ{}^{\ast  \tilde B}+u^{\alpha}_{\beta}(a)da^{\alpha}.
\]
 
We note that the replacement of the coordinates convert the Killing vector field $K_{\alpha}(Q,f)$, the vertical vector field,
\[
K_{\alpha}(Q,f)= K^B_{\alpha}(Q)\frac{\partial}{\partial Q^B}+K^p_{\alpha}\frac{\partial}{\partial { f}^p},
\]
into the vector field $L_{\alpha}=v^{\nu}_{\alpha}(a)\frac{\partial}{\partial a^{\nu}}$ which is   the left-invariant vector field  given on the group manifold $\mathcal G$.
 
The horizontal vector fields are defined with the help of the horizontal projection operators. These operators must extract the direction which is normal to the orbit: ${\Pi}^{\tilde A}_{\tilde E}K^{\tilde E}_{\alpha}=0$. They are defined as follows:
$${\Pi}^{\tilde A}_{\tilde B}={\delta}^{\tilde A}_{\tilde B}-K^{\tilde A}_{\alpha}d^{\alpha \beta}K^{\tilde D}_{\beta}G_{\tilde D \tilde B}.$$
By  ${\Pi}^{\tilde A}_{\tilde B}$, we denote the  four component operator:
$${\Pi}^{\tilde A}_{\tilde B}=({\Pi}^{A}_{B},\, {\Pi}^{A}_{m},\,{\Pi}^{m}_{A},\,{\Pi}^{m}_{n}).$$
The components  are given by the following formulae:
\begin{eqnarray*}
&&{\Pi}^{A}_{B}={\delta}^A_B-K^{A}_{\alpha}d^{\alpha \beta}K^{D}_{\beta}G_{D B}\\
&&{\Pi}^{A}_{m}=-K^{A}_{\mu}d^{\mu \nu}K^{p}_{\nu}G_{pm}\\
&&{\Pi}^{m}_{A}=-K^{m}_{\mu}d^{\mu \nu}K^{D}_{\nu}G_{D A}\\
&&{\Pi}^{m}_{n}={\delta}^m_n-K^{m}_{\mu}d^{\mu \nu}K^{r}_{\nu}G_{rn}.\\
\end{eqnarray*}

The horizontal vector fields are defined as follows:
\begin{equation}
 H_A(Q,f)={\Pi}^R_A\frac{\partial}{\partial Q^R}+{\Pi}^q_A\frac{\partial}{\partial f^q}
\label{H_A(Q)}
\end{equation}
\begin{equation}
 H_p(Q,f)={\Pi}^R_p\frac{\partial}{\partial Q^R}+{\Pi}^m_p\frac{\partial}{\partial f^m}
\label{H_p(Q)}.
\end{equation}
For  $\hat\omega ^{\alpha}$ from (\ref{connect_Q}), we have
\[
 \hat\omega ^{\alpha}(H_A)=0,\;\;\;\hat\omega ^{\alpha}(H_p)=0,\;\;\;\hat\omega ^{\alpha}(K_{\beta})=\delta ^{\alpha}_{\beta}.
\]

Performing the replacement of the coordinate, by means of 
the formulae (\ref{vectfield}), in the expressions (\ref{H_A(Q)}) and (\ref{H_p(Q)}) that represent 
 the horizontal vector fields, we come to the horizontal vector fields
\begin{equation}
H_M(Q^{\ast},\tilde f,a)= \Bigl[N^T_M\Bigl(\frac{\partial}{\partial Q^{\ast T}}-\tilde \mathscr A^{\alpha }_T L_{\alpha}\Bigl)+N^m_M\Bigl(\frac{\partial}{\partial {\tilde f}^m}-\tilde \mathscr A^{\alpha }_mL_{\alpha}\Bigl)\Bigr],
\label{H_A(Q_star)}
\end{equation}
and 
\begin{equation}
 H_m(Q^{\ast},\tilde f,a)=\Bigl( \frac{\partial}{\partial {\tilde f}^m}-\tilde \mathscr A^{\alpha }_mL_{\alpha}\Bigl)
\label{H_p(Q_star)}.
\end{equation}
In  equation (\ref{H_A(Q_star)}), we have used  the components of the projection operator $N^{\tilde A}_{\tilde C}$ :
$$N^{\tilde A}_{\tilde C}=(N^A_C,N^A_m,N^m_A,N^m_p).$$ 
 $N^A_C$ was  defined above. The other components are 
\[
N^A_m=0,\;\;\;  N^m_A=-K^m_{\alpha}({\Phi}^{-1})^{\alpha}_{\mu}\,{\chi}^{\mu}_A=-K^m_{\alpha}{\Lambda}^{\alpha}_A,\;\;\;  N^m_p={\delta}^m_p.
\]
The operator $N^{\tilde A}_{\tilde B}$ satisfy the following properties:
\[
N^{\tilde A}_{\tilde B} N^{\tilde B}_{\tilde C}=N^{\tilde A}_{\tilde C},\;\;\;\;
{\Pi}^{\tilde L}_{\tilde B}N^{\tilde A}_{\tilde L}=N^{\tilde A}_{\tilde B},\;\;\;
{\Pi}_{\tilde L}^{\tilde A}N_{\tilde C}^{\tilde L}={\Pi}_{\tilde C}^{\tilde A}.
\]
Thus, a new coordinate basis consists of 
the horizontal vector fields  (\ref{H_A(Q_star)}) and (\ref{H_p(Q_star)}) together with  the left-invariant vector field $L_{\alpha}$.  

The horizontal coordinate vector fields of this basis do not commute between themselves. They have  the following commutation relations:
\begin{equation}
 [H_A,H_B]={\mathbb C}^T_{AB}\,H_T+{\mathbb C}^p_{AB}\,H_p+{\mathbb C}^{\alpha}_{AB}L_{\alpha}, 
\label{commrelat_AB}
\end{equation}
where the ``structure constants'' are given by 

\[{\mathbb C}^T_{AB}=({\Lambda}^{\gamma}_A N^R_B-{\Lambda}^{\gamma}_BN^R_A) K^{T}_{{\gamma} R},\]

\[{\mathbb C}^p_{AB}=-N^D_AN^R_B({\Lambda}^{\alpha}_{R,D}-{\Lambda}^{\alpha}_{D,R}) K^p_{\alpha}  \;\;  -c^{\sigma}_{\alpha \beta}{\Lambda}^{\beta}_A{\Lambda}^{\alpha}_BK^p_{\sigma}, \]

and
\[{\mathbb C}^{\alpha}_{AB}=-N^S_AN^P_B\,\tilde{\mathcal F}^{\alpha}_{SP}-(N^E_AN^p_B-N^E_BN^p_A)\tilde{\mathcal F}^{\alpha}_{Ep}+N^m_AN^p_B\tilde{\mathcal F}^{\alpha}_{pm}\,.
\]
In ${\mathbb C}^T_{AB}$,  we denote the partial derivative of $K^{T}_{{\gamma}}$ with respect to $Q^{\star}{}^R$ by $K^{T}_{{\gamma} R}$. In ${\mathbb C}^{\alpha}_{AB}$,
the curvature tensor $\tilde{\mathcal F}^{\alpha}_{SP}$ of the connection ${\tilde{\mathscr A}^{\alpha}_P}$ is given by
\[
\tilde{\mathcal F}^{\alpha}_{SP}=\displaystyle\frac{\partial}{\partial Q^{\ast}{}^S}\,\tilde{\mathscr A}^{\alpha}_P- 
\frac{\partial}{\partial {Q^{\ast}}^P}\,\tilde{\mathscr A}^{\alpha}_S
+c^{\alpha}_{\nu\sigma}\, \tilde{\mathscr A}^{\nu}_S\,
\tilde{\mathscr A}^{\sigma}_P,
\]
($\tilde{\mathcal F}^{\alpha}_{SP}({Q^{\ast}},a)={\bar{\rho}}^{\alpha}_{\mu}(a)\,{\mathcal F}^{\mu}_{SP}(Q^{\ast})\,$).   The tensors $\tilde{\mathcal F}^{\alpha}_{Ep}$ and $\tilde{\mathcal F}^{\alpha}_{pm}$ are defined in a similar way.

Next commutation relations are 
\begin{equation}
 [H_A,H_p]={\mathbb C}^m_{Ap}\,H_m+{\mathbb C}^{\alpha}_{Ap}L_{\alpha}
\label{commrelat_Ap}
\end{equation}
with
\[
{\mathbb C}^m_{Ap}=({\bar J}_{\alpha})^m_p{\Lambda}^{\alpha}_A,\;\;\;
{\mathbb C}^{\alpha}_{Ap}=-N^E_A\tilde{\mathcal F}^{\alpha}_{Ep}-N^m_A\tilde{\mathcal F}^{\alpha}_{mp},
 \]
and
\begin{equation}
 [H_p,H_q]={\mathbb C}^{\alpha}_{pq}L_{\alpha}
\label{commrelat_pq}
\end{equation}
with
\[
 {\mathbb C}^{\alpha}_{pq}=-\tilde{\mathcal F}^{\alpha}_{pq}\,.
\]

We notice that the left-invariant vector fields $L_{\alpha}$ of the new basis commute with the coordinate  horizontal vector fields:
\[
 [H_A,L_{\alpha}]=0,\;\;\;[H_p,L_{\alpha}]=0.
\]
Also, for $L_{\alpha}$ we have $[L_{\alpha},L_{\beta}]=c^{\gamma}_{\alpha \beta}L_{\gamma}$.

In a new coordinate basis $(H_A,H_p,L_{\alpha})$,  the  metric tensor (\ref{metric2c})  transforms into the  tensor  
$\check G_{\mathcal A \mathcal B}$ with following components:
\begin{equation}
\displaystyle
{\check G}_{\cal A\cal B}(Q^{\ast},\tilde f,a)=
\left(
\begin{array}{ccc}
{\tilde G}^{\rm H}_{AB} & {\tilde G}^{\rm H}_{Am} & 0\\
{\tilde G}^{\rm H}_{nB} & {\tilde G}^{\rm H}_{nm} & 0\\
0 & 0 &\tilde{d}_{\alpha \beta }  \\
\end{array}
\right)\equiv\left( \begin{array}{cc}
{\tilde G}^{\rm H}_{\tilde A \tilde B}  & 0 \\
0 & \tilde{d}_{\alpha \beta }  \\
\end{array}
\right),
\label{metric2cc}
\end{equation}
where $\tilde{d}_{\alpha \beta }=\rho^{\alpha'}_{\alpha}\rho^{\beta'}_{\beta}d_{\alpha' \beta' }$. 
The  components of the ``horizontal metric'' ${\tilde G}^{\rm H}_{\tilde A \tilde B}$ depending on $(Q^{\ast}{}^A,\tilde f ^m)$ are defined as follows:
$${\tilde G}^{\rm H}_{AB}={\Pi}^{\tilde A}_{A}\,{\Pi}^{\tilde B}_B \,G_{\tilde A\tilde B}=G_{AB}-G_{AD}K^{D}_{\alpha}d^{\alpha \beta}K^R_{\beta}\,G_{RB},$$
because of ${\Pi}^{\tilde C}_{A}\,{\Pi}^{\tilde D}_B \,G_{\tilde C\tilde D}= {\Pi}^{C}_{A}\,{\Pi}^{D}_B \,G_{CD}+
{\Pi}^{q}_{A}\,{\Pi}^{p}_B \,G_{qp}$.

$${\tilde G}^{\rm H}_{Am}=-G_{AB}K^{B}_{\alpha}\,d^{\alpha \beta}K^p_{\beta}G_{pm}.$$
Notice that ${\tilde G}^{\rm H}_{Am}$ is equal to
$${\tilde G}^{\rm H}_{mA}=-G_{mq}K^{q}_{\mu}\,d^{\mu \nu}K^D_{\nu}G_{DA}.$$

$${\tilde G}^{\rm H}_{mn}={\Pi}^r_mG_{rn},\;\;\;\; \rm{or}$$
${\Pi}^{\tilde C}_{m}\,{\Pi}^{\tilde D}_n \,G_{\tilde C\tilde D}= {\Pi}^{C}_{m}\,{\Pi}^{D}_n \,G_{CD}+
{\Pi}^{r}_{m}\,{\Pi}^{q}_n \,G_{rq}=G_{mn}-G_{mr}K^r_{\alpha}d^{\alpha \beta}K_{\beta}^pG_{pn}.$

It worth to note that the metric with components  ${\tilde G}^{\rm H}_{\tilde\mathcal A \tilde\mathcal B}$  is given on the local surface $\tilde \Sigma$ and gives rise the metric on the orbit space $\mathcal P\times _\mathcal G V$, provided that the submanifold  $\tilde \Sigma$ is given parametrically.

The pseudoinverse matrix ${\check G}^{\cal A\cal B}$ to the matrix (\ref{metric2cc}) is represented as
\begin{equation}
\displaystyle
{\check G}^{\cal A\cal B}=
\left(
\begin{array}{ccc}
{G}^{EF}N^A_EN^B_F & {G}^{EF}N^A_EN^q_F & 0\\
{G}^{EF}N^p_F N^B_E & {G}^{pq}+G^{AB}N^p_AN^q_B & 0\\
0 & 0 &\tilde{d}^{\alpha \beta }  \\
\end{array}
\right).
\label{metric2bb}
\end{equation}
This matrix is defined from the following orthogonality condition:
\[
\displaystyle
{\check G}^{\cal A\cal B}{\check G}_{\cal B\cal E}=
\left(
\begin{array}{ccc}
N^A_D& 0 & 0\\
 N^p_D & {\delta}^p_m & 0\\
0 & 0 & {\delta}^{\alpha}_{\beta} \\
\end{array}
\right)\equiv
\left(
\begin{array}{cc}
N^{\tilde A}_{\tilde D} & 0 \\
0 & {\delta}^{\alpha}_{\beta} \\
\end{array}
\right),
\]
where 
\[
\displaystyle
N^{\tilde A}_{\tilde D}=
\left(
\begin{array}{cc}
N^A_D& N^A_m\\
 N^p_D & N^p_m\\
\end{array}
\right)
\]
($N^A_m=0, N^p_m={\delta}^p_m$).

Finally,  it can be shown that  the expression (\ref{lagrang_2}) for the Lagrangian $\mathcal L$ takes the following form in the coordinate basis $(H_A,H_p,L_{\alpha})$:
\begin{equation}
 {\hat{\mathcal L}}=\frac12\,({\tilde G}^{\rm H}_{AB}\, {\omega}^A {\omega}^B +{\tilde G}^{\rm H}_{Ap}\, {\omega}^A {\omega}^p+{\tilde G}^{\rm H}_{pA}\, {\omega}^p {\omega}^A+{\tilde G}^{\rm H}_{pq}\, {\omega}^p {\omega}^q+ {\tilde{d}}_{\mu \nu} {\omega}^{\mu} {\omega}^{\nu})-V,
\label{lagrang_3}
\end{equation}
where we have introduced the new time-dependent variables ${\omega}^A,{\omega}^p$ and ${\omega}^{\alpha}$ that are related to the velocities:
\begin{eqnarray} 
&&\omega ^A=(P_{\bot})^A_B\, \frac{dQ^{\ast}{}^B}{dt}=\frac{dQ^{\ast}{}^A}{dt},\;\;\;\;
\omega ^p=\frac{d\tilde f^p}{dt}\nonumber\\
&&{\omega}^{\alpha}=u^{\alpha}_{\mu}\frac{da^{\mu}}{dt}+{\tilde {\mathscr A}}^{\alpha}_E\,\frac{dQ^{\ast}{}^E}{dt}+{\tilde {\mathscr A}}^{\alpha}_m\, \frac{d\tilde f^m}{dt}.
\label{veloc_omega}
\end{eqnarray}

\section{The Lagrange-Poincar\'{e} equations}

The  Lagrange-Poincar\'{e} equations for the Lagrangian (\ref{lagrang_3}) were obtained  in \cite{preprint} by using the Poincar\'{e} variational principle. They are given by the following equations:
\begin{eqnarray}
&&-\frac{d}{dt}\Bigl(\frac{\partial{\hat{\mathcal L}}}{\partial {\omega}^E}\Bigr)+\Bigl(\frac{\partial{\hat{\mathcal L}}}{\partial {\omega}^T}\Bigr){\mathbb C}^{T}_{CE}\,{\omega}^C+\Bigl(\frac{\partial{\hat{\mathcal L}}}{\partial {\omega}^p}\Bigr)({\mathbb C}^{p}_{CE}\,{\omega}^C+{\mathbb C}^{p}_{qE}\,{\omega}^q)
\nonumber\\
&&\;\;\;\;\;\;\;+\Bigl(\frac{\partial{\hat{\mathcal L}}}{\partial {\omega}^{\alpha}}\Bigr) ({\mathbb C}^{\alpha}_{CE}\,{\omega}^C+{\mathbb C}^{\alpha}_{mE}\,{\omega}^m)+H_E(\hat{\mathcal L})=0,
\label{eq_Poinc_Q}
\end{eqnarray}

\begin{eqnarray}
 &&\!\!\!\!\!\!\!\!\!\!\!\!\!\!\!\!\!\!\!\!\!\!\!\!-\frac{d}{dt}\Bigl(\frac{\partial{\hat{\mathcal L}}}{\partial {\omega}^m}\Bigr)+\Bigl(\frac{\partial{\hat{\mathcal L}}}{\partial {\omega}^p}\Bigr){\mathbb C}^{p}_{Em}\,{\omega}^E
\nonumber\\
&&\!\!\!\!\!\!\!\!\!+\Bigl(\frac{\partial{\hat{\mathcal L}}}{\partial {\omega}^{\alpha}}\Bigr) ({\mathbb C}^{\alpha}_{Em}\,{\omega}^E+{\mathbb C}^{\alpha}_{pm}\,{\omega}^p)+H_m(\hat{\mathcal L})=0,
\label{eq_Poinc_f}
\end{eqnarray}
\begin{equation}
\!\!\!\!\!\!\!\!\!\!\!\!\!\!\!\!\!\!\!\!\!\!\!\!\!\!\!\!\!\!\!\!\!\!\!\!\!-\frac{d}{dt}\Bigl(\frac{\partial{\hat{\mathcal L}}}{\partial {\omega}^{\alpha}}\Bigr)+\Bigl(\frac{\partial{\hat{\mathcal L}}}{\partial {\omega}^{\beta}}\Bigr){c}^{\beta}_{\mu \alpha}\,{\omega}^{\mu}+L_{\alpha}(\hat{\mathcal L})=0.
\label{eq_Poinc_alfa}
\end{equation}
The first two equations of this system are   the horizontal equations, and the last equation, for  the group variable, is the   vertical one.

\section{The  Lagrange-Poincar\'{e} equations in local coordinates}
In this section we show how the resulting Lagrange-Poincar\'{e} equations can be expressed in terms of local coordinates.
First, consider the horizontal equations (\ref{eq_Poinc_Q}) and (\ref{eq_Poinc_f}).

The first term of equation (\ref{eq_Poinc_Q}), a term with a time derivative, can be written as follows:
\[
-\frac{d}{dt}\Bigl({\tilde G}^H_{BT}{\omega}^B +{\tilde G}^H_{pT}{\omega}^p\Bigr)=-\Bigl({\tilde G}^H_{BT}\frac{d{\omega}^B}{dt}+
{\tilde G}^H_{pT}\frac{d{\omega}^p}{dt}+\frac{d}{dt}({\tilde G}^H_{BT}){\omega}^B+\frac{d}{dt}({\tilde G}^H_{pT}){\omega}^p\Bigr),
\]
where
\[
 \frac{d}{dt}({\tilde G}^H_{BT})=(P_{\bot})^M_S\Bigl(\frac{\partial {\tilde G}^H_{BT}}{\partial Q^M}\Bigr)\Bigr|_{Q=Q^{\ast}}{\omega}^S+\frac{\partial {\tilde G}^H_{BT}}{\partial \tilde f^m}{\omega}^m\equiv
{\tilde G}^H_{BT,M}{\omega}^M+{\tilde G}^H_{BT,m}{\omega}^m.
\]
(We recall that $(P_{\bot})^M_S{\omega}^S={\omega}^M$.) Note that the last term $\frac{d}{dt}({\tilde G}^H_{pT}){\omega}^p$ in the above bracket can be similarly represented.

The last term $H_E(\hat{\mathcal L})$ of the equation (\ref{eq_Poinc_Q}) is given by
\begin{eqnarray*}
&&H_E(\hat{\mathcal L})=\frac12N^D_E\Bigr[\frac{\partial {\tilde G}^H_{AB}}{\partial {Q^{\ast D}}}{\omega}^A{\omega}^B
+2\frac{\partial {\tilde G}^H_{pA}}{\partial Q^{\ast D}}{\omega}^p{\omega}^A
+\frac{\partial{\tilde G}^H_{pq}}{\partial Q^{\ast D}}{\omega}^p{\omega}^q
+\frac{\partial {\tilde d}_{\mu \nu}}{\partial Q^{\ast D}}{\omega}^{\mu}{\omega}^{\nu}\Bigr]
\nonumber\\
&&+\frac12N^p_E\Bigr[\frac{\partial {\tilde G}^H_{AB}}{\partial {\tilde f^p}}{\omega}^A{\omega}^B
+2N^p_E\frac{\partial {\tilde G}^H_{nA}}{\partial \tilde f^p}{\omega}^n{\omega}^A
+\frac{\partial {\tilde G}^H_{nq}}{\partial \tilde f^p}{\omega}^n{\omega}^q
+\frac{\partial {\tilde d}_{\mu \nu}}{\partial \tilde f^p}{\omega}^{\mu}{\omega}^{\nu}\Bigr]
\nonumber\\
&&-\frac12N^D_E{\tilde \mathscr A}^{\alpha}_DL_{\alpha}({\tilde d}_{\mu \nu}) {\omega}^{\mu}{\omega}^{\nu} 
-\frac12N^p_E{\tilde \mathscr A}^{\beta}_pL_{\beta}({\tilde d}_{\mu \nu}) {\omega}^{\mu}{\omega}^{\nu}
-N^D_E\frac{\partial V}{\partial Q^{\ast D}}-N^p_E\frac{\partial V}{\partial \tilde f^p}.
\nonumber\\
\end{eqnarray*}
It can be rewritten in the following form:
\begin{eqnarray*}
&&H_E(\hat{\mathcal L})=\frac12N^D_E\Bigr[ {\tilde G}^H_{AB,D}{\omega}^A{\omega}^B
+2{\tilde G}^H_{pA,D}{\omega}^p{\omega}^A
+{\tilde G}^H_{pq,D}{\omega}^p{\omega}^q
+ {\tilde d}_{\mu \nu,D}\,{\omega}^{\mu}{\omega}^{\nu}\Bigr]
\nonumber\\
&&\;\;\;\;+\frac12N^p_E\Bigr[ {\tilde G}^H_{AB,p}{\omega}^A{\omega}^B
+2N^p_E {\tilde G}^H_{nA,p}{\omega}^n{\omega}^A
+ {\tilde G}^H_{nq,p}{\omega}^n{\omega}^q
+ {\tilde d}_{\mu \nu,p}\,{\omega}^{\mu}{\omega}^{\nu}\Bigr]
\nonumber\\
&&\;\;\;\;-\frac12N^D_E{\tilde \mathscr A}^{\alpha}_DL_{\alpha}({\tilde d}_{\mu \nu}) {\omega}^{\mu}{\omega}^{\nu} 
-\frac12N^p_E{\tilde \mathscr A}^{\beta}_pL_{\beta}({\tilde d}_{\mu \nu}) {\omega}^{\mu}{\omega}^{\nu}-N^D_E V_{,D}-N^p_E V_{,p}.
\nonumber\\
\end{eqnarray*}
(Here we have used the  following property satisfied by our  projection operators: $N^M_E (P_{\bot})^D_M=N^D_E$.)

An analogous term of the second equation (\ref{eq_Poinc_f}) is given by 
\begin{eqnarray*}
H_m(\hat{\mathcal L})&=&
\frac12\Bigr[
{\tilde G}^H_{AB,m}{\omega}^A{\omega}^B
+2{\tilde G}^H_{pA,m}{\omega}^p{\omega}^A
+{\tilde G}^H_{pq,m}{\omega}^p{\omega}^q
+{\tilde d}_{\mu \nu,m}{\omega}^{\mu}{\omega}^{\nu}\Bigr]
\nonumber\\
 &&\;\;\;\;\;\;-\frac12{\tilde \mathscr A}^{\beta}_mL_{\beta}({\tilde d}_{\mu \nu}) {\omega}^{\mu}{\omega}^{\nu}- V_{,m}.\nonumber\\
\end{eqnarray*}

For the following it is convenient to represent the horizontal Lagrange\hyph{Poincar\'{e} equations as a system of two equations. They can be written in the matrix form:
\[
 \left(\matrix{{\tilde G}^H_{TB}&{\tilde G}^H_{Tp}\cr{\tilde G}^H_{mB}&{\tilde G}^H_{mp}\cr}\right)
\left(\matrix{\frac{d{\omega}^B}{dt}\cr \frac{d{\omega}^p}{dt}\cr}\right)-\left(\matrix{A_T\cr B_m\cr}\right)=0,
\]
where by $A_T$ and $B_m$ we denote the potential terms  of the first and second equations and as well as  terms that are quadratic in pseudo-velocities.
Multiplying this matrix equation (from the left) by the matrix
\[
 \left(\matrix{{\tilde G}^{EF}N^A_EN^T_F&{\tilde G}^{EF}N^A_EN^m_F\cr {\tilde G}^{EF}N^r_FN^T_E & {\tilde G}^{rm}+{\tilde G}^{FE}N^r_FN^m_E\cr}\right)
\]
we get
\[
 \left(\matrix{N^A_B & 0\cr N^r_B &{\delta}^r_p\cr}\right)
\left(\matrix{\frac{d{\omega}^B}{dt}\cr \frac{d{\omega}^p}{dt}\cr}\right)-\left(\matrix{ A'^A\cr {B'}^r\cr}\right)=0
\]
($N^r_p={\delta}^r_p$). Thus, we have two equations:

\begin{equation}
 N^A_B\frac{d{\omega}^B}{dt}-{\tilde G}^{EF}N^A_E\,(N^T_F\cdot A_T+N^m_F\cdot B_m)=0
\label{eq1}
\end{equation}
and
\begin{equation}
 N^r_B\frac{d{\omega}^B}{dt}+\frac{d{\omega}^r}{dt}-{\tilde G}^{EF}N^r_E(N^T_F\cdot A_T+N^m_F\cdot B_m)-{\tilde G}^{rm}\cdot B_m =0.
\label{eq2}
\end{equation}

Our goal is to obtain a standard coordinate representation for these equations. This can be achieved by means of combining and rearranging the terms of equations, and  
  will  be done as follows.   First   we consider those  terms of the equations  that depend on the quasi-velocities $\omega ^A$ and $\omega ^p$. And then, in the next step, we get the terms that depend on  the group velocities  $\omega ^{\alpha}$. In addition, at the end of our consideration, a remark will be made about the transformation of the potential terms of these equations.

We begin by studying the expression $N^T_FA_T$. 
Those  terms of $A_T$, that are of interest to us, are given by 
\[
-{\tilde G}^H_{BT,M} {\omega}^B{\omega}^M-{\tilde G}^H_{BT,q} {\omega}^B{\omega}^q-{\tilde G}^H_{pT,M} {\omega}^p{\omega}^M-{\tilde G}^H_{pT,q} {\omega}^p{\omega}^q
\]
\[
 +\frac12N^D_T({\tilde G}^H_{AB,D} {\omega}^A{\omega}^B+2{\tilde G}^H_{pA,D} {\omega}^p{\omega}^A+{\tilde G}^H_{pq,D}) {\omega}^p{\omega}^q
\]
\[
 +\frac12N^p_T({\tilde G}^H_{AB,p} {\omega}^A{\omega}^B+2{\tilde G}^H_{nA,p} {\omega}^n{\omega}^A+{\tilde G}^H_{nq,p} {\omega}^n{\omega}^q)
\]
\[
 +({\tilde G}^H_{BM} {\omega}^B+{\tilde G}^H_{pM} {\omega}^p){\mathbb C}^M_{CT}{\omega}^C+({\tilde G}^H_{pA} {\omega}^A+{\tilde G}^H_{pq} {\omega}^q)({\mathbb C}^p_{MT}{\omega}^M+{\mathbb C}^p_{nT}{\omega}^n).
\]
They are multiplied by the projector $N^T_F$.
First we note that  $N^T_FN^p_T=0$. Consequently, after this multiplication, the third line of the preceding expression does not contribute to the equation.
 We note also that $N^T_F\,{\mathbb C}^M_{CT}\,{\omega}^C$=0. 
This follows from an explicit representation  for  ${\mathbb C}^M_{CT}$:
\[
 {\mathbb C}^M_{CT}=({\Lambda}^{\gamma}_CN^R_T-{\Lambda}^{\gamma}_TN^R_C)K^M_{\gamma ,R},
\]
and due to the  properties:
${\Lambda}^{\gamma}_C{\omega}^C=0$ and $N^T_F{\Lambda}^{\gamma}_T$=0.

The last  properties also lead to $N^T_F{\mathbb C}^p_{nT}=0$, since ${\mathbb C}^p_{nT}=-({\bar J}_{\alpha})^p_n {\Lambda}^{\alpha}_T$.
And for $N^T_F{\mathbb C}^p_{MT}$, we have 
\[
N^T_F{\mathbb C}^p_{MT}= -N^R_FN^D_M({\Lambda}^{\alpha}_{R,D}-{\Lambda}^{\alpha}_{D,R})K^p_{\alpha}.
\]
The vanishing of this term can be shown as follows. Taking the partial differential of 
\[
 {\Lambda}^{\alpha}_{R}(Q^{\ast})=({\Phi}^{-1}){}^{\alpha}_{\beta}(Q^{\ast})\,{\chi}^{\beta}_R(Q^{\ast})
\]
with respect to dependent variable $Q^{\ast D}$, we get 
\[
 (P_{\bot})^S_D{\Lambda}^{\alpha}_{R,S}(Q^{\ast})=(P_{\bot})^S_D(\,({\Phi}^{-1}){}^{\alpha}_{\beta,S}\, {\chi}^{\beta}_R+({\Phi}^{-1}){}^{\alpha}_{\beta}\,{\chi}^{\beta}_{R,S}).
\]
(The appearence of $(P_{\bot})^S_D$ in this formula is due to our rule used for differentiation of the functions  with dependent variables.)
If we multiply the obtained formula by $N^D_M$, then, because of 
 $N^D_M(P_{\bot})^S_D=N^S_M$, we come to
\[
 N^S_M{\Lambda}^{\alpha}_{R,S}=N^S_M(\,({\Phi}^{-1}){}^{\alpha}_{\beta,S}\, {\chi}^{\beta}_R+({\Phi}^{-1}){}^{\alpha}_{\beta}\,{\chi}^{\beta}_{R,S}).
\]
 Multiplying this expression by $N^R_F$ and taking into account  the following property: $N^R_F{\chi}^{\beta}_R=0$, we get the  first term of the above representation for  $N^T_F{\mathbb C}^p_{MT}$: 
 \[
 N^R_FN^S_M{\Lambda}^{\alpha}_{R,S}=N^S_MN^R_F\,({\Phi}^{-1}){}^{\alpha}_{\beta}\,{\chi}^{\beta}_{R,S}.
\]
The expression for the second term of  $N^T_F{\mathbb C}^p_{MT}$ can be derived in a similar way. As a result of differentiation, we obtain
\[
 N^R_F{\Lambda}^{\alpha}_{D,R}=N^R_F(({\Phi}^{-1}){}^{\alpha}_{\beta,R}\, {\chi}^{\beta}_D+({\Phi}^{-1}){}^{\alpha}_{\beta}\,{\chi}^{\beta}_{D,R}),
\]
which  should  now be multiplied by $N^D_M$ to get
\[
N^D_M N^R_F{\Lambda}^{\alpha}_{D,R}=N^D_MN^R_F\,({\Phi}^{-1}){}^{\alpha}_{\beta}\,{\chi}^{\beta}_{D,R}.
\]
Hence $N^T_F{\mathbb C}^p_{MT}$ is equal to the difference of  two obtained expressions:
\[
 N^R_F N^D_M {\Lambda}^{\alpha}_{R,D} - N^D_M N^R_F {\Lambda}^{\alpha}_{D,R}=N^D_MN^R_F({\Phi}^{-1}){}^{\alpha}_{\beta}({\chi}^{\beta}_{R,D}-{\chi}^{\beta}_{D,R}).
\]
Because of  ${\chi}^{\beta}_{R,D}={\chi}^{\beta}_{D,R}$, where
$ {\chi}^{\beta}_{R,D}(Q^{\ast})\equiv \frac{\partial ^2 {\chi}^{\beta}(Q)}{\partial Q^R \partial Q^D}\Big |_{Q=Q^{\ast}}$, the right-hand side of the preceding expression is zero.

Thus, those terms in $N^T_FA_T$ that depend on 
$\omega ^A$ and $\omega ^{p}$ are given as follows:

\begin{eqnarray*}
&&-\frac12N^D_F({\tilde G}^H_{BD,M}+{\tilde G}^H_{MD,B}-{\tilde G}^H_{BM,D}){\omega}^B{\omega}^M
\nonumber\\
&&-N^T_F({\tilde G}^H_{BT,q} +{\tilde G}^H_{qT,B}-{\tilde G}^H_{qB,T}){\omega}^q{\omega}^B-
\frac12N^T_F({\tilde G}^H_{pT,q}+{\tilde G}^H_{qT,p}-{\tilde G}^H_{pq,T}) {\omega}^p{\omega}^q.
\nonumber\\
\end{eqnarray*}
This expression can be written in the following form:
\begin{equation}
 -N^T_F\,{}^{ \mathrm  H}{\tilde \Gamma}_{BMT}{\omega}^B{\omega}^M-2N^T_F\,{}^{ \mathrm  H}{\tilde \Gamma}_{qBT}{\omega}^q{\omega}^B-N^T_F\,{}^{ \mathrm  H}{\tilde \Gamma}_{pqT}{\omega}^p{\omega}^q,
\label{N^T_FA_T}
\end{equation}
in which  ${}^{ \mathrm  H}{\tilde \Gamma}_{BMD}$ is define as
\[
{}^{ \mathrm  H}{\tilde \Gamma}_{BMD}\equiv\frac12({\tilde G}^{\rm H}_{BD,M}+{\tilde G}^{\rm H}_{MD,B}-{\tilde G}^{\rm H}_{BM,D}).
\]
And ${}^{ \mathrm  H}{\tilde \Gamma}_{qBT}$ and ${}^{ \mathrm  H}{\tilde \Gamma}_{pqT}$ have an analogous  definitions.

In $B_m$, the terms that depend on $\omega ^A$ and $\omega ^{p}$ are given by 
\[
-{\tilde G}^H_{Bm,R}\, {\omega}^B{\omega}^R-{\tilde G}^H_{Bm,p} {\omega}^B{\omega}^p-{\tilde G}^H_{pm,R} {\omega}^p{\omega}^R-{\tilde G}^H_{pm,q} {\omega}^p{\omega}^q
\] 
\[
 +{\tilde G}^H_{Bq}\, {\omega}^B{\mathbb C}^q_{Em}{\omega}^E+{\tilde G}^H_{pq}\, {\omega}^p{\mathbb C}^q_{Em}{\omega}^E
\]
\[
 +\frac12{\tilde G}^H_{AB,m}\, {\omega}^A{\omega}^B+{\tilde G}^H_{pA,m} {\omega}^p{\omega}^A+\frac12{\tilde G}^H_{pq,m} {\omega}^p{\omega}^q.
\]
Here  
${\mathbb C}^q_{Em}=({\bar J}_{\alpha})^q_m {\Lambda}^{\alpha}_E$. And  from ${\Lambda}^{\alpha}_E
{\omega}^E=0$, it follows that  ${\mathbb C}^q_{Em}{\omega}^E=0$.

Hence for terms in $N^m_FB_m$, we have
 \begin{eqnarray*}
&&\!\!\!\!\!\!\!\!\!\!\!\!-\frac12N^m_F({\tilde G}^H_{Bm,A}+{\tilde G}^H_{Am,B}-{\tilde G}^H_{AB,m}){\omega}^A{\omega}^B
-N^m_F({\tilde G}^H_{Bm,p}+{\tilde G}^H_{pm,B}-{\tilde G}^H_{pB,m}) {\omega}^p{\omega}^B
\nonumber\\
&&\!\!\!\!\!-\frac12N^m_F({\tilde G}^H_{pm,q}+{\tilde G}^H_{qm,p}-{\tilde G}^H_{pq,m}) {\omega}^p{\omega}^q.
\nonumber\\
\end{eqnarray*}
This expression can be rewritten as
\begin{equation}
  -N^m_F\,{}^{ \mathrm  H}{\tilde \Gamma}_{ABm}{\omega}^A{\omega}^B-2N^m_F\,{}^{ \mathrm  H}{\tilde \Gamma}_{pBm}{\omega}^p{\omega}^B-N^m_F\,{}^{ \mathrm  H}{\tilde \Gamma}_{pqm}{\omega}^p{\omega}^q
\label{N^m_FB_m}.
\end{equation}

Summing (\ref{N^T_FA_T}) and (\ref{N^m_FB_m}), we obtain the terms that belong to $N^T_F A_T+N^m_F B_m$:
\[
 -N^{\tilde T}_F\,{}^{ \mathrm  H}{\tilde \Gamma}_{BM\tilde T}{\omega}^B{\omega}^M-2N^{\tilde T}_F\,{}^{ \mathrm  H}{\tilde \Gamma}_{qB\tilde T}{\omega}^q{\omega}^B-N^{\tilde T}_F\,{}^{ \mathrm  H}{\tilde \Gamma}_{pq\tilde T}{\omega}^p{\omega}^q.
\]
Here we have used  the condensed notation for indices, by which  summation over the repeated index $\tilde T$ means that we have two summation: one is taken for $T$, and the other is performed over   some repeated  index, for which we use a small Latin letter. That is, in our case, for example,   $\tilde T\equiv(T,m)$.

Multiplying the resulting expression by $-{\tilde G}^{EF}N^A_E$, one can obtain the following terms of the first Lagrange-Poincaré equations:
\[
 {\tilde G}^{EF}N^A_EN^{\tilde T}_F\Bigl({}^{ \mathrm  H}{\tilde \Gamma}_{BM\tilde T}{\omega}^B{\omega}^M+2\,{}^{ \mathrm  H}{\tilde \Gamma}_{qB\tilde T}{\omega}^q{\omega}^B+{}^{ \mathrm  H}{\tilde \Gamma}_{pq\tilde T}{\omega}^p{\omega}^q\Bigr).
\]

We may also  introduce the Christoffel symbols ${}^{ \mathrm  H}{\tilde \Gamma}_{BM}^{\tilde R}$, ${}^{ \mathrm  H}{\tilde \Gamma}_{qB}^{\tilde R}$ and ${}^{ \mathrm  H}{\tilde \Gamma}_{pq}^{\tilde R}$ for the horizontal (degenerate) metric ${\tilde G}^H_{\tilde R \tilde T}$. They are  defined 
by means   of the  equalities:
\[
 {}^{ \mathrm  H}{\tilde \Gamma}_{BM\tilde T}={\tilde G}^H_{\tilde R \tilde T}{}^{ \mathrm  H}{\tilde \Gamma}_{BM}^{\tilde R},\;\;\;{}^{ \mathrm  H}{\tilde \Gamma}_{qB\tilde T}={\tilde G}^H_{\tilde R \tilde T}{}^{ \mathrm  H}{\tilde \Gamma}_{qB}^{\tilde R}\;\;\;{\rm and}\;\;{}^{ \mathrm  H}{\tilde \Gamma}_{pq\tilde T}={\tilde G}^H_{\tilde R \tilde T}{}^{ \mathrm  H}{\tilde \Gamma}_{pq}^{\tilde R}.
\]
Then, taking into account the following properties  of the projectors: 
\[
N^{\tilde T}_F{\tilde G}^H_{\tilde R \tilde T}={\tilde G}^H_{F\tilde R }, \;\;{\tilde G}^{EF}{\tilde G}^H_{F\tilde R }={\Pi}^E_{\tilde R}, \;\; N^A_E{\Pi}^E_{\tilde R}=N^A_{ R},\;\; ( N^A_E{\Pi}^E_{r}=0),
\]
it becomes possible to rewrite the obtained expression for the discussed terms of the first 
horizontal Lagrange-Poincaré equations as
\begin{equation}
 N^A_R\Bigl({}^{ \mathrm  H}{\tilde \Gamma}^R_{BM}{\omega}^B{\omega}^M+2\,{}^{ \mathrm  H}{\tilde \Gamma}^R_{qB}{\omega}^q{\omega}^B+{}^{ \mathrm  H}{\tilde \Gamma}^R_{pq}{\omega}^p{\omega}^q\Bigr)\equiv N^A_R\,{}^{ \mathrm  H}{\tilde \Gamma}^R_{\tilde B\tilde M}{\omega}^{\tilde B}{\omega}^{\tilde M}.
\label{it1_o_B,o_M}
\end{equation}

In the second horizontal Lagrange-Poincar\'{e} equation, we are interested in by those terms of  
\[
-{\tilde G}^{EF}N^r_E(N^T_F\cdot A_T+N^m_F\cdot B_m)-{\tilde G}^{rm}\cdot B_m 
\]
that depend on $\omega ^A$ and $\omega ^p$. Here we  proceed in the same way as in the case of the first horizontal equation. As a result, we arrive at
\begin{equation}
 (N^r_E\Pi^E_{\tilde R}+\Pi^r_{\tilde R}) \,{}^{ \mathrm  H}{\tilde \Gamma}^{\tilde R}_{\tilde B\tilde M}{\omega}^{\tilde B}{\omega}^{\tilde M}\equiv N^r_{\tilde R}{}^{ \mathrm  H}{\tilde \Gamma}^{\tilde R}_{\tilde B\tilde M}{\omega}^{\tilde B}{\omega}^{\tilde M}.
\label{it2_o_B,o_M}
\end{equation}
(Note that $N^r_p=\delta ^r_p$.)

Now we will consider the terms of the first horizontal Lagrange-Poincar\'{e} equations, which depend on $\omega ^{\mu}$. We begin by studying such terms in $ N ^ T_FA_T$. In $A_T$, they are represented by the following expression:
\begin{eqnarray*}
 &&H_T(\hat{\mathcal L})({\omega}^{\mu})+\Bigl(\frac{\partial \hat{\mathcal L}}{\partial \omega ^{\epsilon}}\Bigr)({\mathbb C}^{\epsilon}_{CT}{\omega}^C
+{\mathbb C}^{\epsilon}_{pT}{\omega}^p)=
\nonumber\\
&&\;\;\frac12\Bigl[N^D_T\frac{\partial {\tilde d}_{\mu \nu}}{\partial Q^{\ast D}}
+N^p_T\frac{\partial {\tilde d}_{\mu \nu}}{\partial {\tilde f}^p}
-N^D_T{\tilde \mathscr A}^{\alpha}_DL_{\alpha}({\tilde d}_{\mu \nu})
-N^p_T{\tilde \mathscr A}^{\beta}_pL_{\beta}({\tilde d}_{\mu \nu})\Bigr]{\omega}^{\mu}{\omega}^{\nu}
\nonumber\\
&&\;\;\;+ {\tilde d}_{\mu \epsilon}\,{\omega}^{\mu}({\mathbb C}^{\epsilon}_{CT}{\omega}^C
+{\mathbb C}^{\epsilon}_{pT}{\omega}^p),
\end{eqnarray*}
in which by $H_T(\hat{\mathcal L})({\omega}^{\mu})$ we denote the ${\omega}^{\mu}$-dependent part of $H_T(\hat{\mathcal L})$.
Multiplying this expression by $N^T_F$ and discarding the corresponding terms due to  $N^T_FN^p_T=0$, we get
\begin{eqnarray*}
&&\frac12N^D_F\Bigl[\frac{\partial {\tilde d}_{\mu \nu}}{\partial Q^{\ast D}}
-{\tilde \mathscr A}^{\alpha}_DL_{\alpha}({\tilde d}_{\mu \nu})
\Bigr]{\omega}^{\mu}{\omega}^{\nu}
+ {\tilde d}_{\mu \epsilon}\,{\omega}^{\mu}({\mathbb C}^{\epsilon}_{CT}{\omega}^C
+{\mathbb C}^{\epsilon}_{pT}{\omega}^p)N^T_F=
\nonumber\\
&&\frac12\Bigl[N^D_F\, {\tilde \mathscr D}_D{\tilde d}_{\mu \nu}\Bigr]{\omega}^{\mu}{\omega}^{\nu}
+ N^T_F{\tilde d}_{\mu \epsilon}\,{\omega}^{\mu}({\mathbb C}^{\epsilon}_{CT}{\omega}^C
+{\mathbb C}^{\epsilon}_{pT}{\omega}^p), 
\nonumber\\
\end{eqnarray*} 
where the covariant derivative is defined as
\[
N^D_F\, ({\tilde \mathscr D}_D{\tilde d}_{\mu \nu})=  N^D_F\Bigl[\frac{\partial} {\partial Q^{\ast D}} {\tilde d}_{\mu \nu}
-c^{\sigma}_{\alpha \mu}{\tilde \mathscr A}^{\alpha}_D{\tilde d}_{\sigma \nu}
-c^{\sigma}_{\alpha \nu}{\tilde \mathscr A}^{\alpha}_D{\tilde d}_{\sigma \mu}
\Bigr],
\]
$
{\tilde d}_{\mu \nu}={\rho}^{\mu '}_{\mu}(a) {\rho}^{\nu '}_{\nu}(a)d_{\mu' \nu'}(Q^{\ast},\tilde f)
$ and ${\tilde {\mathscr A}}^{\alpha}_D={\bar \rho}^{\alpha}_{\mu}(a){\mathscr A}^{\mu}_D(Q^{\ast},\tilde f)$.

Recalling that
\[
 {\mathbb C}^{\epsilon}_{pT}=-N^R_T{\tilde {\mathscr F}}^{\epsilon}_{pR}-N^m_T{\tilde \mathscr F}^{\epsilon}_{pm},
\]
\[
 {\mathbb C}^{\epsilon}_{CT}=-N^R_CN^Q_T{\tilde \mathscr F}^{\epsilon}_{RQ}-(N^R_CN^p_T-N^R_TN^p_C){\tilde \mathscr F}^{\epsilon}_{Rp}
+N^m_CN^p_T{\tilde \mathscr F}^{\epsilon}_{pm},
\]
and using the following equalities: $N^R_C{\omega}^C={\omega}^R$, $N^p_C{\omega}^C=-K^p_{\mu}{\Lambda}^{\mu}_C{\omega}^C=0$, it is not difficult to show that the expression for   $\omega ^{\mu}$-terms in  $N^T_FA_T$ is given by  
\begin{eqnarray*}
&&-N^T_F {\tilde d}_{\mu \epsilon}\,{\omega}^{\mu}({\tilde \mathscr F}^{\epsilon}_{QT}{\omega}^Q+{\tilde \mathscr F}^{\epsilon}_{pT}{\omega}^p)+\frac12\bigl(N^T_F\, {\tilde \mathscr D}_T{\tilde d}_{\mu \nu}\bigr){\omega}^{\mu}{\omega}^{\nu}
\equiv
\nonumber\\
&&-N^T_F{\tilde d}_{\mu \epsilon}{\tilde \mathscr F}^{\epsilon}_{{\tilde Q}T}{\omega}^{\tilde Q}{\omega}^{\mu}+\frac12\bigl(N^T_F\, {\tilde \mathscr D}_T{\tilde d}_{\mu \nu}\bigr){\omega}^{\mu}{\omega}^{\nu}.
\nonumber\\
\end{eqnarray*}

In $B_m$,  $\omega ^{\mu}$-terms are represented by the following expression:
\begin{eqnarray*}
 &&H_m
(\hat{\mathcal L})({\omega}^{\mu})+\Bigl(\frac{\partial \hat{\mathcal L}}{\partial \omega ^{\epsilon}}\Bigr)({\mathbb C}^{\epsilon}_{Em}{\omega}^E
+{\mathbb C}^{\epsilon}_{pm}{\omega}^p)=
\nonumber\\
 &&\frac12{\tilde \mathscr D}_m({\tilde d}_{\mu \nu}){\omega}^{\mu}{\omega}^{\nu}
+ {\tilde d}_{\mu \epsilon}\,{\omega}^{\mu}({\mathbb C}^{\epsilon}_{Em}{\omega}^E
+{\mathbb C}^{\epsilon}_{pm}{\omega}^p),
\nonumber\\
\end{eqnarray*}
in which   the covariant derivative is defined as
$${\tilde \mathscr D}_m({\tilde d}_{\mu \nu})={\partial {\tilde d}_{\mu \nu}}/{\partial {\tilde f}^m}
-c^{\sigma}_{\beta \mu}{\tilde \mathscr A}^{\beta}_m{\tilde d}_{\sigma \nu}
-c^{\sigma}_{\beta \nu}{\tilde \mathscr A}^{\beta}_m{\tilde d}_{\sigma \mu}.$$

Therefore  in   $N^m_FB_m$, they are given by
\begin{eqnarray*}
&&\frac12N^m_F{\tilde \mathscr D}_m({\tilde d}_{\mu \nu}){\omega}^{\mu}{\omega}^{\nu} -N^m_F{\tilde d}_{\mu \epsilon}\,{\omega}^{\mu}({\tilde \mathscr F}^{\epsilon}_{Rm}{\omega}^R+{\tilde \mathscr F}^{\epsilon}_{pm}{\omega}^p)\equiv
\nonumber\\
&&N^m_F\Bigl(\frac12{\tilde \mathscr D}_m({\tilde d}_{\mu \nu}){\omega}^{\mu}{\omega}^{\nu}-{\tilde d}_{\mu \epsilon}{\tilde \mathscr F}^{\epsilon}_{{\tilde Q}m}{\omega}^{\tilde Q}{\omega}^{\mu}\Bigr).
\nonumber\\
\end{eqnarray*}
Hence    $\omega ^{\mu}$-terms  of $N^T_FA_T+N^m_FB_m$ are as follows:
\begin{equation}
 -N^{\tilde R}_F{\tilde d}_{\mu \epsilon}{\tilde {\mathscr F}}^{\epsilon}_{\tilde Q\tilde R}\,{\omega}^{\tilde Q}{\omega}^{\mu}+\frac12N^{\tilde R}_F({\tilde \mathscr D}_{\tilde R}{\tilde d}_{\mu \nu}){\omega}^{\mu}{\omega}^{\nu}.
\label{omeg_NTAT+NmBm}
\end{equation}

The obtained expressions  (\ref{it1_o_B,o_M}), (\ref{it2_o_B,o_M}) and (\ref{omeg_NTAT+NmBm}), when used them in  the equations (\ref{eq1}) and (\ref{eq2}), lead to  new coordinate representations of the horizontal Lagrange-Poincar\'{e} equations:
 \begin{eqnarray*}
&&N^A_B\frac{d{\omega}^B}{dt}+N^A_R\,{}^{\rm  H}\tilde {\Gamma}^R_{\tilde B\tilde M}{\omega}^{\tilde B} {\omega}^{\tilde M} +
\nonumber\\
&&\;\;\;\;\;\;\;\;\;\;\;\;\;\;\;G^{EF}N^A_EN^{\tilde R}_F\Bigl[{\tilde d}_{\mu\epsilon}{\tilde \mathscr F}^{\epsilon}_{\tilde Q\tilde R}  {\omega}^{\tilde Q} {\omega}^{\mu} -\frac12(\tilde{\mathscr D}_{\tilde R} {\tilde d}_{\mu\nu}) {\omega}^{\mu}{\omega}^{\nu}+V_{\!,\tilde R}\Bigr]=0                                 
 \end{eqnarray*}
and
\begin{eqnarray*}
&&\!\!\!N^r_B\frac{d{\omega}^B}{dt}+\frac{d{\omega}^r}{dt}+N^r_{\tilde R}\,{}^{\rm  H}\tilde {\Gamma}^{\tilde R}_{\tilde A\tilde B}{\omega}^{\tilde A} {\omega}^{\tilde B}+G^{EF}N^r_FN^{\tilde R}_E\Bigl[{\tilde d}_{\mu\epsilon}{\tilde \mathscr F}^{\epsilon}_{\tilde Q\tilde R}  {\omega}^{\tilde Q} {\omega}^{\mu}-
\nonumber\\
&&\!\!\!\frac12(\tilde{\mathscr D}_{\tilde R} {\tilde d}_{\mu\nu}) {\omega}^{\mu}{\omega}^{\nu}+V_{\!,\tilde R}\Bigr]+ G^{rm}\Bigl[{\tilde d}_{\mu\epsilon}{\tilde \mathscr F}^{\epsilon}_{\tilde Q m}  {\omega}^{\tilde Q} {\omega}^{\mu}-\frac12(\tilde{\mathscr D}_{m} {\tilde d}_{\mu\nu}) {\omega}^{\mu}{\omega}^{\nu}+V_{\!,m}\Bigr]=0.
\nonumber\\
\end{eqnarray*}
We note that because of  the  invariance of the potential $V(Q^{\ast},\tilde f)$ under the action of the group $\mathcal G$,   in these equations, we have, in fact, $N^{\tilde R}_FV_{\!,\tilde R}=V_{\!, F}$.

Our final transformation of these equations is the replacement of the variable $\omega ^{\mu}$ by the new variable $p_{\alpha}=d_{\alpha \mu'}{\rho}^{\mu'}_{\mu}{\omega}^{\mu}$. 
As a result, we obtain
\begin{eqnarray}
&&\!\!\!\!\!\!\!\!\!N^A_B\frac{d{\omega}^B}{dt}+N^A_R\,{}^{\rm  H}\tilde {\Gamma}^R_{\tilde B\tilde M}{\omega}^{\tilde B} {\omega}^{\tilde M} +
\nonumber\\
&&G^{EF}N^A_EN^{\tilde R}_F\Bigl[{ \mathscr F}^{\alpha}_{\tilde Q\tilde R}  {\omega}^{\tilde Q} p_{\alpha} +\frac12({\mathscr D}_{\tilde R} {d}^{\kappa\sigma}) p_{\kappa}p_{\sigma}+V_{,\tilde R}\Bigr]=0,                                 
\label{itog_hor_1}
\end{eqnarray}
\begin{eqnarray}
&&\!\!\!\!\!\!\!\!\!\!\!\!N^r_B\frac{d{\omega}^B}{dt}+\frac{d{\omega}^r}{dt}+N^r_{\tilde R}\,{}^{\rm  H}\tilde {\Gamma}^{\tilde R}_{\tilde A\tilde B}{\omega}^{\tilde A} {\omega}^{\tilde B}+
G^{EF}N^r_FN^{\tilde R}_E\Bigl[{ \mathscr F}^{\alpha}_{\tilde Q\tilde R}  {\omega}^{\tilde Q} p_{\alpha}+
\nonumber\\
&&\!\!\!\!\!\!\!\!\!\!\!\!\frac12({\mathscr D}_{\tilde R} {d}^{\kappa\sigma}) p_{\kappa}p_{\sigma}+V_{,\tilde R}\Bigr]+
G^{rm}\Bigl[{ \mathscr F}^{\alpha}_{\tilde Q m}  {\omega}^{\tilde Q} p_{\alpha}+\frac12({\mathscr D}_{m} {d}^{\kappa\sigma}) p_{\kappa}p_{\sigma}+V_{,m}\Bigr]=0.
\label{itog_hor_2}
\end{eqnarray}

Now consider the third  Lagrange-Poincar\'{e} equation. 
For the Lagrangian (\ref{lagrang_3}),   this vertical  Lagrange-Poincar\'{e} equation (\ref{eq_Poinc_alfa}) is as follows:
\[
 -\frac{d}{dt}\bigl(\tilde d_{\mu \alpha}\omega ^{\mu}\bigr)+\tilde d_{\mu \epsilon}\omega ^{\mu}c^{\epsilon}_{\nu\alpha}\omega ^{\nu}+\frac12L_{\alpha}\bigl(\tilde d_{\mu \nu}\bigr)\omega ^{\mu}\omega ^{\nu}=0.
\]
Since
\[
 \frac12L_{\alpha}\bigl(\tilde d_{\mu \nu}\bigr)=\frac12L_{\alpha}\bigl(\rho^{\mu'}_{\mu}\rho^{\nu'}_{\nu}\bigr) d_{\mu' \nu'}=\frac12(c^{\gamma}_{\alpha \mu}\rho^{\mu'}_{\gamma}\rho^{\nu'}_{\nu}+c^{\gamma}_{\alpha \nu}\rho^{\nu'}_{\gamma}\rho^{\mu'}_{\mu})d_{\mu' \nu'}, 
\]
we have 
\[
 \frac12L_{\alpha}\bigl(\tilde d_{\mu \nu}\bigr)=\frac12(c^{\gamma}_{\alpha \mu}\tilde d_{\gamma \nu}+c^{\gamma}_{\alpha \nu}\tilde d_{\gamma \mu}).
\]
This means that the second and third terms of the Lagrange-Poincar\'{e} equation cancel each other out, and we get
\[
 -\frac{d}{dt}\bigl(\tilde d_{\mu \alpha}\omega ^{\mu}\bigr)=0.
\]
The resulting equation, resembling the conservation law  of the "color charge", can be rewritten in terms of the dual variable 
$p_{\alpha}$, which  we  introduced  above.

Replacing ${\omega}^{\mu}$ by $p_{\alpha}=d_{\alpha \mu'}{\rho}^{\mu'}_{\nu}{\omega}^{\nu}$, we  obtain\[
 \frac{d}{dt}(\rho ^{\nu}_{\alpha} p_{\nu})\equiv\rho ^{\nu}_{\alpha}\frac{d}{dt}p_{\nu}+\frac{\partial \rho ^{\nu}_{\alpha}}{\partial a^{\mu}}\frac{d a^{\mu}}{dt}p_{\nu}=0.
\]
But since
\[
 \frac{d a^{\beta}}{dt}= v^{\beta}_{\alpha}(\omega ^{\alpha}-\omega ^E{\tilde \mathscr A}^{\alpha}_E-\omega ^p {\tilde \mathscr A}^{\alpha}_p) \,\,\,\,\,\rm{with}\,\,\,{\tilde \mathscr A}^{\alpha}_{\tilde E}=\bar \rho ^{\alpha}_{\sigma}\mathscr A ^{\sigma}_{\tilde E}, 
\]
and $L_{\alpha}\rho ^{\gamma}_{\beta}=c^{\mu}_{\alpha \beta}\rho ^{\gamma}_{\mu}$, 
 after using this substitution it can be shown that the vertical Lagrange-Poincar\'{e} equation is as follows:
\begin{equation}
 \frac{d p_{\beta}}{dt}+ c^{\nu}_{\mu \beta}d^{\mu \sigma}p_{\sigma}p_{\nu}-c^{\nu}_{\sigma \beta}\mathscr A^{\sigma}_{\tilde E}\omega ^{\tilde E}p_{\nu}=0.  
\label{itog_vert}
\end{equation}
Thus, this equation, together with the equations (\ref{itog_hor_1}) and (\ref{itog_hor_2}), are the local Lagrange-Poincar\'{e} equations written in terms of the local coordinates.

\section{The equations for relative equilibrium}
Having derived the local Lagrange-Poincar\'{e} equations, we will make a brief remark about the equation for relative equilibria.

We recall that for mechanical systems with symmetry, the
   relative equilibrium is a special movement of the original system, which in the case of a projection onto the reduced manifold becomes the equilibrium of the reduced mechanical system.  
From the theory of dynamical systems with the symmetry \cite{Marsden} it is known that in a relative equilibrium the system performs the stationary motion. In addition, in this motion the shape of the system does not change. So to get the equations for finding the relative equilibria, we need to put ${\omega}^{\tilde A} = 0$ (${\omega}^A = 0$, ${\omega}^p = 0$) in the Lagrange-Poincar\'{e} equations. The horizontal equations (\ref{itog_hor_1}) and (\ref{itog_hor_2}) become as follows: 
\[
 G^{EF}N^A_EN^{\tilde R}_F\Bigl[\frac12({\mathscr D}_{\tilde R} {d}^{\kappa\sigma}) p_{\kappa}p_{\sigma}+V_{,\tilde R}\Bigr]=0
\]
and
\[
G^{EF}N^r_FN^{\tilde R}_E\Bigl[\frac12({\mathscr D}_{\tilde R} {d}^{\kappa\sigma}) p_{\kappa}p_{\sigma}+V_{,\tilde R}\Bigr]+
 G^{rm}\Bigl[\frac12({\mathscr D}_{m} {d}^{\kappa\sigma}) p_{\kappa}p_{\sigma}+V_{,m}\Bigr]=0.
\]
They can be rewritten as 
\begin{eqnarray*}
\left\{ 
\begin{array}{l}
 \displaystyle
 G^{EF}N^A_EN^{ R}_F\cdot \bigl( 1 \bigr)_R+G^{EF}N^A_EN^{ m}_F\cdot \bigl( 2 \bigr)_m=0\\
 G^{EF}N^{ r}_FN^R_E\cdot \bigl( 1 \bigr)_R+(G^{rm}+G^{EF}N^r_FN^m_E)\cdot \bigl( 2 \bigr)_m=0,
\end{array}\right.
\end{eqnarray*}
where
\[
 \bigl( 1 \bigr)_R=\frac12({\mathscr D}_{R} {d}^{\kappa\sigma}) p_{\kappa}p_{\sigma}+V_{,R}
\]
and
\[
 \bigl( 2 \bigr)_m=\frac12({\mathscr D}_{m} {d}^{\kappa\sigma}) p_{\kappa}p_{\sigma}+V_{,m}.
\]
In the matrix form this system of equations  looks like
\begin{eqnarray*}
 \displaystyle
\left(
 \begin{array}{cc}
  G^{EF}N^A_EN^{ R}_F & G^{EF}N^A_EN^m_F\\
 G^{EF}N^r_FN^R_E & G^{rm}+G^{EF}N^r_FN^m_E\\
\end{array}
\right)
\left(
\begin{array}{c}
 \bigl( 1 \bigr)_R \\
\bigl( 2 \bigr)_m
\end{array}
\right)=0.
\end{eqnarray*}
Multiplying it from the left by the matrix
\[ 
\displaystyle
\left(
\begin{array}{cc}
G^H_{BA} & G^H_{Br}\\
G^H_{pA} & G^H_{pr}\\
\end{array}
\right),
\]
we obtain
\[
  \displaystyle
\left(
 \begin{array}{cc}
  N^R_B & N^m_B\\
 0 & \delta ^m_p\\
\end{array}
\right)
\left(
\begin{array}{c}
 \bigl( 1 \bigr)_R \\
\bigl( 2 \bigr)_m
\end{array}
\right)=0.
\]
That is, we have
\begin{eqnarray*}
\left\{
\begin{array}{l}
\displaystyle
N^R_B\cdot \bigl( 1 \bigr)_R +N^m_B\cdot \bigl( 2 \bigr)_m=0\\
 \;\;\;\;\;\;\;\;\;\;\;\;\;\;\;\;\;\;\;\;\delta ^m_p\cdot \bigl( 2 \bigr)_m=0.
\end{array}\right.
\end{eqnarray*}
But this means that 
\begin{eqnarray*}
\left\{
\begin{array}{l}
\displaystyle
N^R_B\cdot \bigl( 1 \bigr)_R=0\\
 \bigl( 2 \bigr)_m=0.
\end{array}\right.
\end{eqnarray*}
In other words,
\begin{eqnarray}
\left\{
\begin{array}{l}
\displaystyle
N^R_B \Bigl(\frac12({\mathscr D}_{R} {d}^{\kappa\sigma}) p_{\kappa}p_{\sigma}+V_{,R}\Bigr)=0\\
\;\;\; \frac12({\mathscr D}_{m} {d}^{\kappa\sigma}) p_{\kappa}p_{\sigma}+V_{,m}=0.
\end{array}\right.
\label{itog_hor_eqlbr}
\end{eqnarray}
It should  be noted that although these resulting equations  look as if they are independent, but really it is not so. They are interrelated, since the matrix $d^{\mu \sigma}$ is inverse to the matrix representing the sum of two orbital metrics ${\gamma}_{\mu \nu}$ and $ {\gamma}'_{\mu\nu }$.

For solvability of the equations (\ref{itog_hor_eqlbr}) it is required (in the standard approach) that $p_{\alpha}={\rm const}$. 
 Taking this condition into account, the vertical Lagrange-Poincare equation (\ref{itog_vert}) is transformed into
\begin{eqnarray}
c^{\nu}_{\mu \beta}d^{\mu \sigma}p_{\sigma}p_{\nu}=0.
\label{itog_vert_eqlbr}
\end{eqnarray}
Thus, (\ref{itog_hor_eqlbr}) and (\ref{itog_vert_eqlbr}) are the equations for determining the relative equilibria of the  mechanical system under consideration.

Note that (\ref{itog_vert_eqlbr}) can be solved using the  eigenvectors of the matrix $d^{\mu \sigma}$
\cite{Littlejohn}.  
Namely, $p_{\alpha}$ is assumed to be proportional to the eigenvector of this matrix.

\end{document}